\documentclass[10pt,letterpaper]{article}
\usepackage[margin=1.25in]{geometry}

\usepackage[utf8]{inputenc}

\usepackage[authoryear]{natbib}
\setcitestyle{authoryear, open={(},close={)}}
\makeatletter
\renewcommand{\@biblabel}[1]{\quad#1.}
\makeatother

\usepackage{microtype}
\DisableLigatures[f]{encoding = *, family = * }

\usepackage[aboveskip=1pt,labelfont=bf,labelsep=period,singlelinecheck=off]{caption}

\usepackage{color}
\definecolor{Gray}{gray}{.25}

\usepackage{graphicx}
\usepackage{nameref,hyperref}
\usepackage{amsmath}
\usepackage{booktabs}
\usepackage{siunitx}
\usepackage{multirow}
\usepackage{bbm}

\begin{document}
\vspace*{0.35in}

\begin{flushleft}
{\Large
\textbf\newline{Non-coresident family as a driver of migration change in a crisis: \\The case of the COVID-19 pandemic}
}
\newline
\\
Unchitta Kan\textsuperscript{1,*},
Jericho McLeod\textsuperscript{1},
Eduardo L{\'{o}pez}\textsuperscript{1}
\\
\bigskip
\bf{1} Department of Computational and Data Sciences, George Mason University, VA, USA
\\
\bigskip
* ukanjana@gmu.edu

\end{flushleft}

\bigskip
\noindent\textbf{Original submitted draft:}  October 23, 2023\\
\noindent\textbf{Current revision:} April 1, 2024
\bigskip

\section*{Abstract}
Changes in U.S. migration during the COVID-19 pandemic show that many moved to less populated cities from larger cities, deviating from previous trends. In this study, building on prior work in the literature showing that the abundance of family ties is inversely related to population size, we analyze these migration changes with a focus on the crucial, yet overlooked factor of extended family. Employing two large-scale data sets, census microdata and mobile phone GPS relocation data, we show a collection of empirical results that paints a picture of migration change affected by kin. Namely, we find that people migrated closer to family at higher rates after the COVID-19 pandemic started. Moreover, even controlling for factors such as population density and cost of living, we find that changes in net in-migration tended to be larger and positive in cities with larger proportions of people who can be parents to adult children (our proxy for parental family availability, which is also inversely related to population size). Our study advances the demography-disaster nexus and amplifies ongoing literature highlighting the role of broader kinship systems in large-scale socioeconomic phenomena.

\bigskip
\noindent\textbf{Keywords: } inter-city migration, kinship, family distribution, cities, demography, mobility data
\bigskip



\section{Introduction}

The COVID-19 pandemic brought about a profound change in how people live and work. Among other things, the closure of facilities and attractions as well as the shift to remote work weakened the connection between people's place of residence and work. Many individuals were no longer constrained to live in a particular city for economic opportunities, and some were incentivized by financial or personal reasons to relocate. It is not surprising, then, that notable changes in U.S. domestic migration trends were observed after the pandemic started. Our analysis, as well as others~\citep{coven2023jue, haslag2022boise}, suggests that a substantial proportion of these deviations can be attributed to a modification in the destinations of people's migration: more people moved to smaller (less populated) cities, especially from larger cities, and fewer people moved to large cities.

What characteristics of cities of different sizes attracted or repelled people during the pandemic? Less crowding, lower cost of living, measures introduced to control the pandemic, and even weather may have helped to explain the disruption of typical migration patterns and why people were moving to smaller cities. In this study, we investigate a factor that has been overlooked: extended family who reside elsewhere. We ask the question, \emph{can some of these migration changes be attributable to individuals relocating to be closer to kin?}

Our question is motivated by a recent empirical finding~\citep{mcleod2023origins} that an individual is more likely to have kin ties located in smaller cities, as well as lines of research that point to the critical role of kin ties~\citep{david2023society, furstenberg2020kinship, mulder2018putting} and how they are particularly activated during crises~\citep{reed2023communication, volker2023networks}. Importantly, COVID-19 was a major societal event that, at least temporarily, uncoupled geography and economic aspirations—the latter of which is in tension with extended family orientation~\citep{miller1976family}. If family-driven migration was indeed happening at higher rates, flows to smaller cities would follow as a consequence of the distribution of non-coresident family ties in the U.S.~\citep{mcleod2023origins}.

In this paper, we investigate empirically the connection between changes in inter-city migration during the COVID-19 crisis, city population, and the role that non-coresident family played in both large-scale patterns and individual-level migration dynamics. Our analyses show these factors to be related, emphasizing that the distribution of people's non-coresident family is relevant in order to understand human migration. Here, for reasons explained below, our test of the influence of family on migration is for parental relationships. Our results point to important implications in the context of cities, and amplify ongoing literature which highlights the need for more research on the role of broader kinship in large-scale socioeconomic phenomena~\citep{david2019network, david2023society}, especially non-household kin that is outside of the nuclear family~\citep{furstenberg2020kinship, reed2023communication}.


\subsection{Non-coresident family as a driver of migration in a crisis}\label{sec:family-background}

There are at least three lines of research that help us understand how non-household kin may influence migration decisions in normal times, and why it is plausible that people would relocate to be closer to them in times of crisis.
The first line is referred to as the `family ties perspective'~\citep{mulder2007family, mulder2018putting}. The second line is the literature which focuses on how social ties are activated in a crisis such as the COVID-19 pandemic~\citep{reed2023communication, volker2023networks}. The third line is the well-known tension between personal economic and social achievements~\citep{miller1976family}.


\subsubsection{Importance of kin}
The `family ties perspective'~\citep{mulder2007family, mulder2018putting} argues that family ties need to be taken into account in migration analyses because of three crucial ingredients they offer: support, need for proximity, and uniqueness.

Family is central to social and support networks of people~\citep{wellman1979community, dunbar1995social, plickert2007s, rozer2016family}. While individuals can, and do, maintain ties with family via calls after moving away~\citep{david2023society,lambiotte2008geographical}, family face-to-face interaction and transfer of practical or physical support cannot be fully replaced by virtual interaction; they require geographical proximity and thus incentivize migration.

Kin is fundamentally different from other types of relationships because it is given rather than chosen. Also, an important characteristic of kinship that is less prominent in other relationships is the feeling of responsibility that kin feels towards each other~\citep{mulder2018putting}. From transfers of resources to help with mundane tasks to emotional support to aversion of crises, people look to their kin. For example, parents and adult children are more than \emph{ten times} as likely to give or receive major assistance compared to other types of relationships~\citep{wellman1979community}. In 2012 alone, it was estimated that informal care, which includes kin-based intergenerational care, amounted to over one billion hours of unpaid work per week in the U.S.~\citep{dukhovnov2015takes}. People place greater importance on kin than non-kin when it comes to interpersonal contacts~\citep{david2023society}, especially women in their child-rearing years; collectively, individuals may even prioritize family over themselves~\citep{krys2023family}.

Research has also shown that merely having family in one's social network can influence the composition of that network itself~\citep{dunbar1995social,rozer2016family}. At the same time, the composition of one's support network can also become focused on kin ties in non-routine situations.


\subsubsection{Crises and kin ties}

In times of crises, kin ties may be particularly activated. For example, as formal care facilities shut down during the pandemic~\citep{lee2021care}, informal intergenerational care (e.g. day care of young children by grandparents or eldercare by adult children) was likely to be even more important. This may be evidenced in the study by \citeauthor{volker2023networks}~\citeyearpar{volker2023networks}, which indicated that while people's social networks became smaller and focused on core ties after the COVID-19 pandemic started, the network of practical helpers of the elderly proportionally consisted more of their children; similarly, among individuals aged 18–35, parents made up a larger share of their practical helpers networks.

Simultaneously, during the pandemic, \citeauthor{reed2023communication}~\citeyearpar{reed2023communication} found an increase in communications with non-coresident kin. \citeauthor{tunccgencc2023social}~\citeyearpar{tunccgencc2023social} found family bonds to positively influence well-being, and that among close social ties, only family bonds were positively linked to engagement in health behaviors. These findings are in line with another study by~\citeauthor{lee2023transformation}~\citeyearpar{lee2023transformation} showing the stability and strengthening of bonds with kin compared to other ties after the start of the pandemic.

The activation of kin ties in systemic crises is not a phenomenon specific to just COVID-19. Other examples dating decades earlier include \citeauthor{shavit1994kin}~\citeyearpar{shavit1994kin} who have found similar results in the context of the Gulf War, and \citeauthor{hulbert2000core}~\citeyearpar{hulbert2000core} in the context of hurricane Andrew.


\subsubsection{Extended kin in migration decisions} 

While previous literature findings have established that non-household kin can influence migration decisions (see, e.g., ~\citeauthor{kan2007residential}~\citeyearpar{kan2007residential}; \citeauthor{spring2017influence}~\citeyearpar{spring2017influence}), kinship factors may be in tension with economic aspirations in the propensity to migrate~\citep{miller1976family}. In other words, people may have to choose between being closer to economic opportunities or to extended family. However, it can be argued that COVID-19 is a unique case in that it is a systemic crisis that enabled more mobility, not less. At a time where individuals and families might look to their family the most, the pandemic also decoupled geography and employment, allowing people to achieve proximity to their extended family to a greater extent than before.


\subsubsection{Parental ties}\label{sec:parental-ties}

Among kin ties, intergenerational ties are understood in the literature to be the ``important arena of action in Western kinship systems''~\citep{ furstenberg2020kinship}. The majority of support from kin flows vertically (typically in the downward direction, i.e., from parents to children or grandchildren). That about 75\% of people in the U.S. whose parents or children are still alive reside within 30 miles from one of them~\citep{choi2020spatial} reflects this situation. Parents-in-law often act similarly to parents in terms of the support they provide~\citep{wellman1989brothers, compton2014family}, and in many cultures they can be considered consequential in terms of one's social network and cooperation~\citep{david2023network}. While there is some modest evidence of horizontal transfer of time and resources between family members such as siblings~\citep{wellman1989brothers, white2001sibling}, relatively little is known about contacts and exchanges between extended kin such as aunts, uncles, cousins, etc.~\citep{furstenberg2020kinship}.

Aside from the clear evidence of their prominence, parental ties are a very compelling variable to study as a ``pull factor'' in pandemic-migration for the reason that migration is highly age- and life course-specific as well as context-dependent~\citep{millington2000migration}. 

Individuals aged between 18 and 44 tend to have the highest propensity to migrate~\citep{rogerson2005population, molloy2011internal}. This propensity declines with age~\citep{castro1984age}, which may have to do with the accumulation of social capital~\citep{kan2007residential}; it also declines with the household family life cycle stage~\citep{miller1976family}. 

At the same time, individuals in the 18 – 44 age range may also be the most likely to need support from or give support to parental family. For example, dual-earner parents returning to live near their own parents for childcare assistance is an identified phenomenon in the literature~\citep{bailey2004migration}. Grandparents are known to be important providers of early childcare~\citep{furstenberg2020kinship, dukhovnov2015takes, mulder2018putting}, and evidence suggests proximity to them increases labor force participation among mothers with young children~\citep{compton2014family}. In the other care direction, individuals in their late child-rearing period—who belong to the so-called ``sandwich generation''—may also find themselves needing to care for their aging parents. 

Therefore, if it is largely the individuals in their twenties, thirties, and forties that were migrating for familial reasons, it is likely that they would look to where their parents were. For this reason, we focus in this study primarily on parental family.


\section{Materials and methods}


\subsection{Design of study}\label{sec:design}

Our study design seeks to provide evidence for the existence of a three-way relation between pandemic-migration, city population size, and non-coresident family. First, we analyze large-scale relocation data and show the increased migration flow to cities with smaller population after the pandemic started. Then, through three empirical investigations, we link migration to parental family as well as parental family to population—the first is done at the individual level, while the other two are done at the city level.

In Investigation 1, we examine whether there are differences in the individual-level migration rates to move towards parental family (or ``return to home'') after the pandemic began in 2020. In Investigation 2, we construct a proxy variable that estimates the abundance of people who can be parents to adult children in each U.S. city and relate it to both city population and migration variables. In Investigation 3, we estimate an empirical model which tests whether cities with higher parental family availability saw a higher net population influx after the pandemic started.

Overall, Investigation 1 serves to validate our line of inquiry (linking migrational change to kin) as well as our study assumptions, while Investigations 2 and 3 connect parental family availability to both city population and changes in net migration of cities. The empirical model controls for relevant migration factors and mitigates potential confounders. Put together, our study examines whether the trend to migrate to smaller cities is partially a consequence of people moving to be closer to family, coupled with the heterogenous spatial distribution of parental family ties in the U.S.

We now elaborate on these analyses and the methodology we employ, beginning with the data used in this study.


\subsection{Data sets}\label{sec:data}

The two primary, large-scale data sets we employ in this study are the county-to-county relocation index aggregated from anonymized, opted-in mobile phone GPS data provided by the location intelligence company~\citet{spectus}, and the yearly U.S. Census American Community Survey Public Use Microdata Samples (ACS PUMS).

The Spectus data set has an advantage over administrative place-level migration data (e.g., from the U.S. Internal Revenue Service) due to its higher temporal resolution (weekly) as well as its ability to capture real time migration by algorithmically detecting new home location from mobile phone GPS instead of relying on individuals to report their change of address to governmental agencies (which could occur with a significant delay or not occur at all). Our data encompass nearly one-fourth of all possible origin-destination city pairs in the U.S.

Our second large-scale data set, ACS PUMS, contains individual-level information on a subsample of the U.S. Census ACS respondents, including the demographics of each person in the household, their place of birth, and their current place of residence. The microdata sample also includes survey questions about migration in the past year, such as whether a respondent has moved and the location from which they moved. For our study, we obtain PUMS yearly samples for the years 2016 to 2021 (corresponding to six separate samples) from IPUMS USA~\citep{ruggles2023integrated} which integrates and harmonizes PUMS data across all survey years. Across the six years, our IPUMS USA sample contains in total $N=18,694,272$ person-records from $N=8,247,978$ households.

Supplementarily, we obtain city population sizes, as well as control variables used in our empirical model from the ACS aggregated estimates~\citep{censusacs} and the~\citet{bea-employment}.


\subsection{Calculations of inter-city migration from Spectus data}\label{sec:methods-spectus}

In the raw form, the Spectus data measures the weekly relocation flows between pairs of counties in the U.S. as an aggregated index called the Relocation Index. To calculate the Relocation Index $r_{hk}(t)$ during week $t$ between county $h$ (old county) and county $k$ (new county), Spectus uses the following formula: 
\begin{equation}
    r_{hk}(t) = \frac{\text{devices in $h$ with a new home in $k$ in week $t$}}{\text{total number of devices in $h$}}.
\end{equation}
Home location is detected using an algorithm that identifies persistent night-time GPS location. For movers, this algorithm detects a change in the home location. Note that the Spectus data preserves the privacy of the users as they are aggregated at the county level. Our data spans from January 2019 to December 2020.

Our study is focused on inter-city migration, and we take core-based statistical areas (CBSAs) to be the geographic unit of study. CBSAs are urban areas delineated by the U.S. Office of Management and Budget, and can be either micropolitan areas (with population between 10,000 and 49,999 people) or metropolitan areas (with population of at least 50,000 people). We estimate dyadic relocation flows at the CBSA level by performing additional aggregations on the county-level Spectus data. In particular, to estimate the number $R_{ij}(t)$ of moves (as opposed to an index) from city $i$ to city $j$ during week $t$, we use the formula 
\begin{equation}\label{eq:R}
    R_{ij}(t) = \sum_{h \in i}\sum_{k \in j} \pi_h p_h r_{hk}(t),
\end{equation}
where the sums are over the counties that have a geographic correspondence to the appropriate CBSAs (i.e., counties that are part of the CBSAs), $p_h$ is the population of county $h$, and $\pi_h$ is the estimated device sampling rate for county $h$. We calibrated $\pi_h$ by calculating and comparing yearly Spectus flow at the county level to the average yearly county-to-county migration flow derived from the 5-year, 2016–2019 migration data published by the~\citet{census-migration}. 

We also calculate the number $R_{ij}(\theta)$ of relocations from city $i$ to city $j$ spanning over a certain time period as
\begin{equation}\label{eq:Rtheta}
    R_{ij}(\theta) = \sum_{t\in\theta} R_{ij}(t) =  \sum_{t\in\theta}\sum_{h \in i}\sum_{k \in j} \pi_h p_h r_{hk}(t).
\end{equation}
We use $\theta=0$ to indicate the period between April 2019 – December 2019, which we take to be the baseline period in this study, and $\theta=1$ to indicate the period between April 2020 – December 2020, the pandemic comparison period. Because we are only concerned with inter-city migration and not intra-city flows, we set $R_{ii}(t)$ and $R_{ii}(\theta)$ to 0.

In total, there are 211,902 origin-destination city pairs for the baseline period and 192,946 origin-destination pairs for the comparison period; both periods comprise 926 unique origin CBSAs and 926 unique destination CBSAs. The smallest city in the data set is Lamesa, TX (with population around 13,000 people) and the largest city is the New York metropolitan area (with population around 19.3 million people).


\subsection{Analysis of trend to move to smaller cities during COVID-19}\label{sec:methods-trend}

To illustrate the trend to move to smaller cities during COVID-19—the phenomenon which motivated this study—we define $z(P'; P, \theta)$, the probability that movers from origin CBSAs whose log-population sizes fall into the log-population bin $P+\Delta P$ would relocate to a destination CBSA whose log-population falls into the $P'+\Delta P$ bin during period $\theta$ (where $\Delta P$ is the bin size). Binning is employed to ease interpretation of the results and to manage fluctuations from the sparsity of samples of city population sizes, and we use log-population instead of raw population due to the skewed nature of city sizes in the U.S.~\citep{IOANNIDES}. Using the migration quantity derived from the Spectus data (equation~\ref{eq:Rtheta}), we calculate $z(P'; P, \theta)$ as follows:
\begin{equation}
    z(P'; P,\theta)=\frac{\sum_{i\in P}\sum_{j\in P'}R_{ij}(\theta)}{\sum_{i\in P}\sum_{j}R_{ij}(\theta)},
    \label{eq:z}
\end{equation}
where the notation $\sum_{i\in P}$ indicates summation over cities $i$ whose log-population sizes belong to a bin $P+\Delta P$. 

To capture the changes in $z$ after the pandemic started, we calculate the ratio $z(P'; P,\theta=1)/z(P';P,\theta=0)$ for all $P', P$.


\subsection{Investigation 1: Micro-level analysis of return-to-home movers}\label{sec:methods-micro}

\begin{table}[t!]
\centering
\begin{tabular}{c | c | c | c } 
Migration rate $\lambda_m(t)$ & Migration type & Who moved & Where moved  \\
\hline
$\lambda_1(t)\quad(m=1)$ & Type 1 & Individual movers & parents' home \\
$\lambda_2(t)\quad(m=2)$ & Type 2 & Individual movers & native place\\
\; & \; & \; & (not joining parents') \\
$\lambda_3(t)\quad(m=3)$ & Type 3 & Family household movers & native place
\end{tabular}
\caption{\color{Gray} Types of moves identified in the ACS PUMS microdata, analyzed across the years 2016–2021. These types of moves represent relocations towards parental and extended family ties (native place is used in the literature as a proxy for presence of family ties~\citep{compton2014family}). Details and discussion are given in Sections~\ref{sec:methods-micro} and~\ref{sec:results-micro}.}
\label{table:pums-move-types}
\end{table}

As a first investigation to examine potential links between non-coresident family and migration change during the COVID-19 pandemic, we employ the IPUMS USA microdata samples to study and compare the micro-level behavior of movers across the years 2016 to 2021. We examine whether there are differences in migration rates to move towards parental family after the pandemic began in 2020. In this exercise, we do not yet relate migration to specific cities or city sizes due to limitations in the data (namely, the PUMS variables of interest are only available at the levels of U.S. state and Census Public Use Microdata Areas). However, this exercise validates our line of inquiry (linking migrational change to kin) as well as study assumptions that will be relevant to our next investigation.

We classify three types of family-driven, ``return-to-home'' movers in the IPUMS USA data to analyze their patterns: (1) individuals moving into their parents' households; (2) individuals moving back to their native state from elsewhere but not joining their parents' household; and (3) family household units moving back to their native state from elsewhere. We call these types of moves Type 1, 2, and 3, respectively (see Table~\ref{table:pums-move-types}). 

Type 1 movers are relatively simple to interpret: these are individuals who moved back in with their parents. 
For some adults and households that are a family unit, however, moving in with their parents may not be an option. Instead, they may choose to relocate to be within the locality of their parents. Precise information about the residence of one's parents is highly identifiable and not publicly available. Therefore, we follow the literature~\citep{compton2014family} and use one's native state as a proxy variable for proximity to or presence of family ties. The residence of one's parents is often the home in which one grew up, which is in turn often located in one's state of birth. We refer to such returns to place of nativity performed by individuals as Type 2 moves, and we refer to such moves performed by family household units as Type 3 moves.

As a technical point which applies to both Type 1 and 2 movers, we consider individual movers to be persons who had migrated in the past year at the time of the PUMS survey, who were either a household of one person or a person who resided in a household in which not everyone had moved. If a household were labeled in the survey as a family household, were larger than one person, and every householder has moved, then we consider the household to be a family household unit mover (which is relevant to identifying Type 3 movers).

To identify Type 1 movers, we look at individual movers who had migrated within the same state or between states in the past year at the time of the survey (i.e., whose `\texttt{MIGRATE1}' variable values are either 2 or 3 in the encoding of the IPUMS USA data). For each person in the sample, IPUMS USA includes variables that identify the mother (`\texttt{MOMLOC}') and father (`\texttt{POPLOC}') of that person if they live in the same household. (These variables are calculated probabilistically by IPUMS USA as they are not present in the regular ACS PUMS.) For each individual mover, if at least one parent was present in their current household and the parent(s) had not also moved in the past year, we classify that individual as a Type 1 mover.

To identify Type 2 movers, we look at IPUMS USA individual movers who had migrated from a different state in the past year at the time of the survey  (`\texttt{MIGRATE1}' = 2). If their current state of residence (`\texttt{STATEFIP}') was the same as their state of birth (`\texttt{BPL}'), we label them as Type 2 movers (i.e., individual movers moving back to place of nativity from elsewhere). We note that, strictly speaking, Type 1 moves are not mutually exclusive from Type 2 because one's parental household may be in one's native state (so Type 2 moves may contain some Type 1 moves). However, for our study, we exclude Type 1 moves from Type 2 moves.

Finally, we classify family households (`\texttt{HHTYPE}' is 1, 2, or 3) larger than one person whose every member had migrated in the past year to be Type 3 movers if their migration destination (current state of residence) was the place of birth of at least one householder.

We introduce the quantity $\lambda_m(t)$, where $m \in \{1,2,3\}$ indexes the type of movers, to capture the rates that their respective types of migration occur in each year $t$. Table~\ref{table:pums-move-types} provides a summary (who and where moved) of these move types and their corresponding $\lambda_m(t)$. For notational brevity, we assume the dependence on $t$ in $\lambda_m(t)$ is implicit and write $\lambda_m$ interchangeably.  We calculate $\lambda_1$ and $\lambda_2$ for each year by dividing the number of Type 1 and Type 2 movers, respectively, by the total number of individual movers. We calculate $\lambda_3$ by dividing the number of Type 3 movers by the total number of household movers. Our calculations are with consideration to sampling weights (see Section 4 of the publication by~\citeauthor{censusacs-doc}~\citeyearpar{censusacs-doc} for ACS PUMS sampling weights). 
Explicitly, we use the formula
\begin{equation}
    \lambda_m(t) = \frac{\sum_{u\in s(t)} w_u \mathbbm{1}_{\text{ind}}(u) \mathbbm{1}_m(u)}{\sum_{u \in s(t)} w_u \mathbbm{1}_{\text{ind}}(u)} \quad\quad\quad (m=1,2)
\end{equation}
to calculate $\lambda_1$ and $\lambda_2$, where $u$ denotes persons in the IPUMS USA sample $s(t)$ in year $t$, $\mathbbm{1}_{\text{ind}}$ and $\mathbbm{1}_m$ are binary variables ($=1$ if a condition is met and 0 if it is not) indicating whether $u$ is an individual mover and Type $m$ mover, respectively. The variable $w_u$ corresponds to the person sampling weight (`\texttt{PERWT}') of $u$. 

For $\lambda_3$, we use
\begin{equation}
    \lambda_3(t) = \frac{\sum_{q\in s(t)} w_q \mathbbm{1}_{\text{hh}}(q) \mathbbm{1}_3(q)}{\sum_{q \in s(t)} w_q \mathbbm{1}_{\text{hh}}(q)},
\end{equation}
where $q$ denotes households in the yearly IPUMS USA sample, $\mathbbm{1}_{\text{hh}}$ and $\mathbbm{1}_3$ are binary variables indicating whether $q$ is a family household mover and Type 3 mover, respectively, and $w_q$ corresponds to the household sampling weight (`\texttt{HHWT}') of $q$. 

We note that for the 2020 sample, IPUMS USA uses experimental weights published by U.S. Census Bureau to address data collection and quality issues associated with the COVID-19 pandemic. Although the results for 2020 should still be interpreted with caution, we include them because they nevertheless provide valuable information.


\subsection{Investigation 2: City-level analysis of parental family availability, population size, and net migration}\label{sec:methods-macro}

If people were to migrate back to parental family after the COVID-19 shock, then we should see also larger flows to cities which have a larger abundance of parental family households. \citet{mcleod2023origins} refer to this abundance as \emph{availability}. For our study, we require measures of parental family availability and net migration at the city level, described in this section. 


\subsubsection{Constructing parental family availability proxy variable from IPUMS USA}

As mentioned earlier, there is considerable scarcity of data linking people to the location of their parents. Therefore, in this exercise we introduce a proxy variable $v_i$ which estimates the stock of households in each city $i$ whose householder(s) can be parents to adult children, capturing the notion of parental family availability. 

Motivated by the literature and by methodological reasons, we design our proxy variable $v_i$ as follows. Because PUMS data do not contain information on kin who do not reside in the same household as the sample individuals, we cannot directly infer the city in which an individual's parents may be located. At the same time, based on the literature we discussed in Section~\ref{sec:parental-ties}~\citep{miller1976family, castro1984age, millington2000migration, rogerson2005population, kan2007residential, molloy2011internal, mulder2018putting},
we assume that relocating individuals tend to generally be in their twenties, thirties, or forties, because these demographic groups tend to have the highest propensities to migrate. (We validate this assumption using statistics of the age of ``return-to-home'' movers in our previous analysis.) With this assumption, we expect that their parents would be at least a generation older. Therefore, we base our parental household estimation on certain age and marriage criteria. 

Explicitly, within the IPUMS USA 2019 sample, we identify households in the sample that satisfy the following criteria:
\begin{enumerate}
    \item Family households in which the head of household or their spouse (if married and spouse is present) is at least 50 years old, or
    \item Non-family households in which the head of household is at least 50 years old and is either married but no spouse present, separated, divorced, or widowed.
\end{enumerate}
The specific age of 50 was chosen based on the literature finding that ``the vast majority of American parents who are older than the age of 50 provide support to children and grandchildren'' \citep{furstenberg2020kinship}. Summing the household sampling weights (`\texttt{HHWT}') of these households estimates the total number of such households in the sample.

The most detailed geographic identifier in the ACS PUMS is the Public Use Microdata Area (PUMA) which consists of one or more contiguous counties and census tracts. Because our goal is to analyze migration patterns at the city level, we perform a PUMA-to-CBSA geo-allocation mapping algorithm to obtain city estimates of the parental family proxy variable. The mapping algorithm relies on the geographic correspondence file between PUMAs and CBSAs obtained from the Geocorr application maintained by the~\citet{geocorr}. In this correspondence file, each entry is a PUMA–CBSA intersection along with an allocation factor which represents the proportion of the population living in this intersection out of the entire PUMA. Using these factors, we allocate the weighted total number of households satisfying the above criteria in each PUMA to each CBSA that intersects with it. Finally, we divide this number in each CBSA by the weighted total number of households in the CBSA to obtain the share of parental family households $v_i$.


\subsubsection{City-level net migration}

From our Spectus inter-city flows $R_{ij}(t)$ and $R_{ij}(\theta)$, defined in equations~(\ref{eq:R}) and~(\ref{eq:Rtheta}) respectively, we derive two \emph{net} migration quantities: $y_i(t)$, and $y_i(\theta)$.

The quantity $y_i(t)$ is a measure of net in-migration of a city and captures the inflow per outflow of city $i$ during week $t$. It is given by
\begin{equation}\label{eq:yit}
    y_{i}(t) = \frac{\sum_{k} R_{ki}(t)}{\sum_{j} R_{ij}(t)},
\end{equation}
where the numerator gives the total flow into a city $i$ from all other cities during week $t$ and the denominator gives the total flow out of a city $i$ to all other cities during week $t$. Notice that $y_{i}(t) > 1$ implies that the migration inflow exceeds the migration outflow of city $i$ during week $t$ (i.e., positive net influx). Consequently, $y_{i}(t)$ captures the direction and magnitude of the net migration flow of a city.

We also calculate the corresponding quantity spanning the comparison time period, $y_{i}(\theta)$, using
\begin{equation}\label{eq:yitheta}
    y_{i}(\theta) = \frac{\sum_{k} R_{ki}(\theta)}{\sum_{j} R_{ij}(\theta)}.
\end{equation}

Similar to the approach in Section~\ref{sec:methods-trend}, we can calculate the ratio $y_i(\theta = 1)/y_i(\theta=0)$ to capture the changes in the net in-migration between the two time periods. We relate both $y_i(t)$ and $y_i(\theta=1)/y_i(\theta=0)$ to our parental family proxy $v_i$.


\subsection{Investigation 3: Empirical model}\label{sec:methods-model}
Finally, we estimate an empirical model to help control for other factors that may have been at play in pandemic-migration. We use a difference-in-differences (DiD) strategy with a continuous treatment variable (namely, $v_i$). DiD is an econometric model used to estimate the effect of a treatment by comparing the outcomes in the treated and untreated groups between two time periods~\citep{lee2016matching}; a continuous treatment is used to model increasing intensity of treatment instead of splitting units into treated and untreated groups. Applied to our study, with our outcome being changes in net in-migration, we can model parental family availability as a continuous treatment and take the two time periods to be before and after the COVID-19 shock.

Our model, which pools data from both before and during the pandemic, can be written as
\begin{equation}
    \ln \left[ \frac{y_{i}(\theta=1)}{y_{i}(\theta=0)} \right] = \beta v_i + \gamma + \sum_a \rho_a C_{ai},
\label{eq:regression}
\end{equation}
where the dependent variable is the log-ratio of the net flux into a city (see equation~(\ref{eq:yitheta})) after and before the pandemic started, measuring the changes in migration. $C$ denotes scalar control variables, indexed by $a$. Notice that the city-specific differences that existed in the dependent variable across cities before the COVID-19 shock are accounted for by the denominator in the log-ratio, constituting the city fixed effect.

Our coefficient of interest, $\beta$, is city-dependent and measures whether higher parental family availability $v$ would lead to higher net flux after the pandemic started. 
The coefficient $\gamma$ is city-independent and accounts for the effects that COVID-19 alone had on the dependent variable, constituting the time fixed effect. Finally, the city-dependent control variables $C_a$ are included because we cannot ignore the possibility that the changes in migration behavior may have also been influenced by other factors whose importance may vary after the pandemic started. Using population density as an example, individuals may experience a higher desire to move to less dense places during the pandemic to avoid being infected, in which case the $\rho$ coefficient for population density will be negative. We include control variables that are relevant to relocation decisions: population size and density, median home value; median income; employment level (number of jobs per person); and the share of single family homes (SFH) in the city (home ownership is a major aspiration in the American life, and with work-from-home policies, households may have had more flexibility to seek a location where SFH were more available). All control variables except the share of SFH are in natural log scale.


\section{Results}


\subsection{Increased migration from large to small cities}\label{sec:results-pattern}

\begin{figure}[t!]
    \centering
    \includegraphics[width=\textwidth]{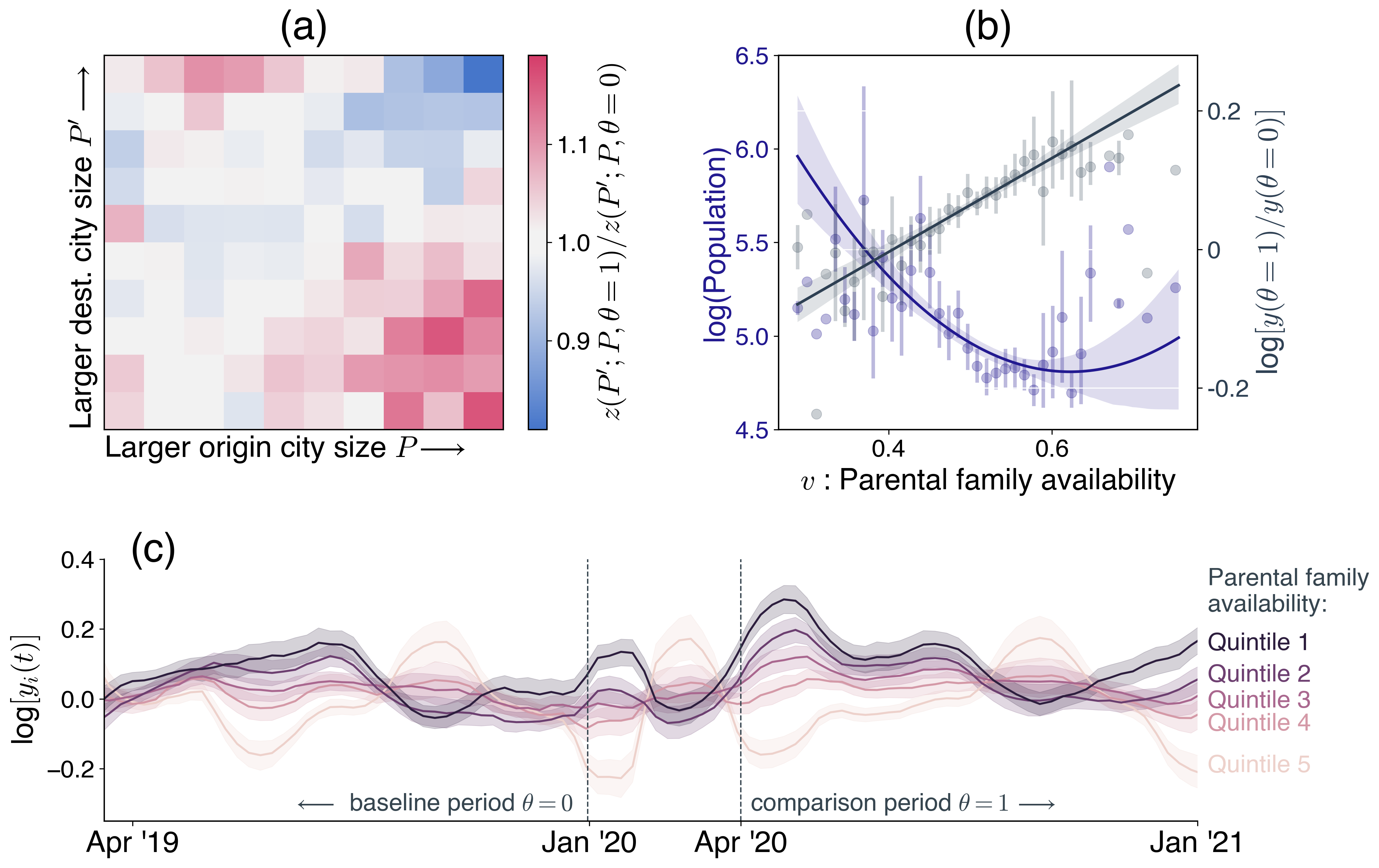}
    \caption{\color{Gray} Changes in relocation patterns after the COVID-19 pandemic started in terms of city population size and parental family availability. Panel (a) shows the changes in the probability $z$ to relocate between (binned) city sizes before and during COVID-19 and suggests that movers from large cities were much more likely to relocate to small cities after the pandemic started than during the 2019 baseline period (red region in the bottom right corner). Panel (b) shows a binned scatterplot of our parental family availability proxy $v$ in relation to log-population (blue, left vertical axis) and net in-migration changes (grey, right vertical axis) after COVID-19 shock. The dots represent the mean vertical axis values given the horizontal axis bins (with 50 discrete, equidistant bins along the $v$-axis in total); error bars represent the 95\% CI of the means. The grey line and blue curves are fitted regression lines for the means (linear and order-2 polynomial, respectively) with the shaded regions corresponding to the 95\% CI of the regression estimates. The grey vertical axis is $\log \left[ y(\theta=1) / y(\theta=0) \right]$, which measures how much more (or less) of an attractor the cities in each bin became after the pandemic started (larger positive values indicate that on average the cities saw larger inflow per outflow after the pandemic started). Panel (c) shows a time-series of the average inflow per outflow of cities (log scale) grouped by quintiles of parental family availability $v$ (shaded regions correspond to the 95\% CI of the means), suggesting proportionally high increases in net in-migration to cities in the high $v$-quintiles after the COVID-19 shock in April 2020 compared to the corresponding time in the prior year.}
    \label{fig:1}
\end{figure}

We first present the changes in the probability to relocate to cities with log-population bin $P'$ from an origin city with log-population bin $P$. In Fig.~1, we visualize the ratio $z(P';P,\theta=1)/z(P';P,\theta=0)$ as a function of $P$ and $P'$  (see Section~\ref{sec:methods-trend} and equation~(\ref{eq:z})). We select $b=10$ equisized population bins to achieve granularity without significant sparsity (however, we find qualitatively consistent results for $b=5,...,10$). Fig.~1a reveals that there is a considerable increase in the probability for movers from large U.S. cities to migrate to small cities after the pandemic started (red region in the bottom right corner of Fig.~1a). At the same time, people from large cities were also less likely to migrate to another large city.

In numbers, our estimates derived from the Spectus data (see Section~\ref{sec:methods-spectus}) indicate cities that were smaller than $500,000$ in population had an influx of almost $52,000$ people from cities that had over $500,000$ in population \textit{in excess} of what was observed during the baseline period in 2019. In total, between April 2020 – December 2020, these cities saw an increase in their \emph{net} in-migration by 80 percent as compared to the same period in 2019 ($60,000$ versus $100,000$). In other words, the excess influx from cities larger than 500,000 in population accounts for 95\% of the increase in the net in-migration to these cities whose population is smaller than $500,000$. Meanwhile, the top 10 largest cities saw twice as much net out-migration between April 2020 and December 2020 as compared to the same period in 2019 ($-82,000$ versus $-45,600$).

If it is the case that part of this migration change is due to by family-driven migration, the increased migration to smaller cities may have followed as a consequence of the uneven distribution of family availability across the U.S.~\citep{mcleod2023origins}, where people in larger cities are less likely to have non-coresident family living nearby (hence would migrate elsewhere towards family). And because of the population distribution of cities in the U.S., in which the total number of people living across all cities with population size in a bin around $P$ decays with $P$~\citep{IOANNIDES}, there is a bias in the direction of more extended family being located in cities with progressively smaller $P$. To see if this is a valid line of inquiry (i.e., to check the premise), we determine if individuals or households made the decision to ``move back home'' to be closer to family at a higher rate once the pandemic started in comparison to the pre-pandemic period.


\subsection{Micro-level dynamics of ``moving back home''}\label{sec:results-micro}

\begin{figure}[t]
    \centering
    \includegraphics[width=\textwidth]{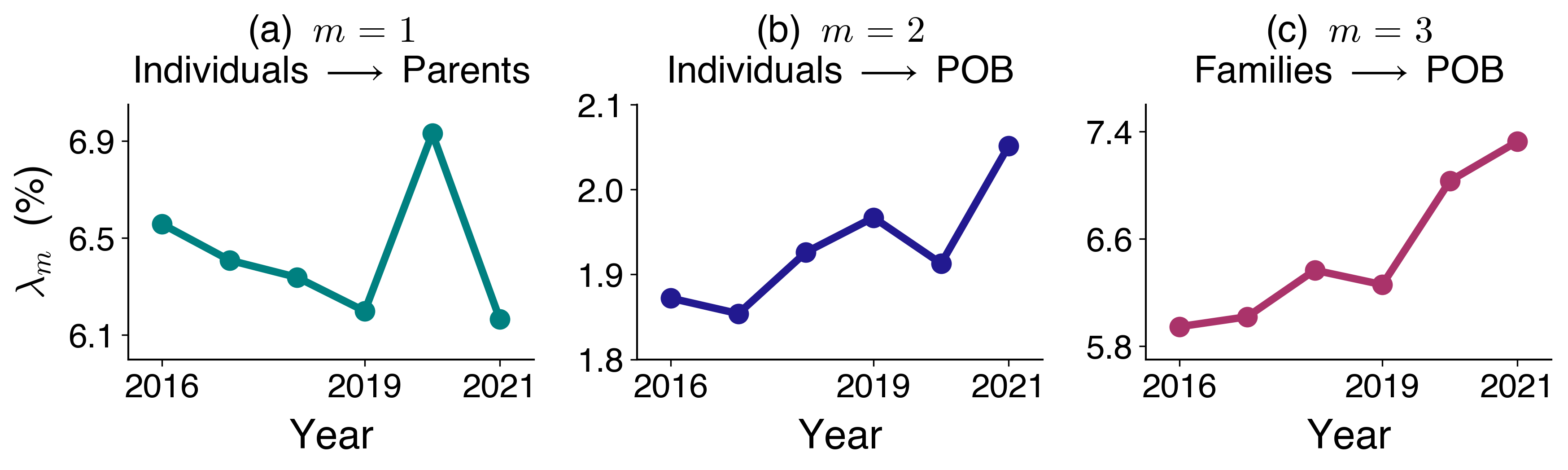}
    \caption{\color{Gray} Results for the micro-level analysis of movers in the IPUMS data, showing the rates for individual movers to (a) move into their parents' households ($\lambda_1$, Type 1 movers in the microdata) and (b) move back to their place of birth (POB) from elsewhere but not joining their parents' household ($\lambda_2$, Type 2 movers). Panel (c) shows $\lambda_3$, the rates for Type 3 moves, i.e., family household units moving back to their native place from elsewhere.}
    \label{fig:2}
\end{figure}

We analyze the rates $\lambda_m(t)$, where $m \in \{1,2,3\}$, at which the three types of return-to-home migration we classified in the IPUMS USA microdata occurred in each year $t$ between 2016 and 2021 (for methodology, see Investigation 1, Section~\ref{sec:methods-micro}). Table~\ref{table:pums-move-types} provides a summary of these move types and their corresponding $\lambda_m(t)$. 
If parental households became a more important ``insurance'' destination under the COVID-19 crisis, we should expect to see a spike in $\lambda_1$ in the year 2020. Similarly, we should see increases in $\lambda_2$ and $\lambda_3$ after 2019.

In Fig.~2a, we indeed observe a jump in the Type 1 migration rate $\lambda_1$ in 2020. However, this rate dropped back to pre-pandemic levels in 2021. A possible interpretation of this result is that individuals who were able to (and perhaps needed to) move back in with their parents did so promptly after the pandemic started. 

Analyzing the demographics of Type 1 movers in our IPUMS USA samples, we find that they tend to be young (median age of 25) and have low income (mean income of \$25,266). More than half of them did not have a college degree, and about 60 percent were employed. Interestingly, the mean and median income of movers, as well as the percentage of college degree holders, were higher in 2020 and 2021 than in pre-pandemic years.

On the other hand, $\lambda_2$, capturing the rate for Type 2 moves in which individuals moved to place of birth from elsewhere, saw a decrease in 2020 but an increase in 2021 (Fig.~2b). While not attempting to provide an explanation, a possible interpretation is that those individuals who did not have the option to move in with parents waited until 2021 to move back to their native state. Compared to Type 1 movers, Type 2 movers tended to be older (median age 29) and with a higher mean income. Similar to Type 1 movers, the mean and median income of Type 2 movers, as well as the proportion of college degree holders, were higher in 2020 and 2021 than in pre-pandemic years.

In Fig.~2c, we observe an increase in $\lambda_3$ in 2020, and even a larger one in 2021, indicating that the pattern for family households to move back to their native states from elsewhere (Type 3 moves) increased in prevalence in 2020 and continued to do so in 2021. Plausibly, relocating an entire family household requires more logistical planning and ``wait-and-see'', which could help explain the continued increase in the rate to move back to native place. 

An interesting temporal-demographic dynamic we observe is that households that performed Type 3 moves in 2021 tended to have slightly older householders compared to prior years. Moreover, compared to prior years, a smaller proportion of family movers in 2020 and 2021 had eldest children who were younger than 5 or between 5 to 10 years old, but a slightly larger proportion of them had eldest children who were in their teens. In other words, families with young children were less likely to perform a move to native state during the pandemic, whereas those with older children (which would also indicate parents who are at the end of their child-rearing years) perform such moves at a marginally higher rate. This provides a modest support for the possibility that eldercare was a more prominent direction of care for the ``sandwich generation'' during the pandemic and, perhaps, a more important driver of family-related migration at the time.

Finally, we note that the demographics of movers here conform to our expectations (that they tend be inbetween their twenties and forties), which also helps to validate study assumptions that we rely on when constructing our family proxy variable $v$ in our next analysis.


\subsection{Cities with higher parental family availability observed larger positive changes in net in-migration}\label{sec:results-large-scale}

Up to now, we have shown that people moved more to smaller places and that moves towards place of origin also increased. In this section, we show in more detail that the parental family availability variable $v_i$ is both negatively correlated with population size and positively correlated with an increase in net in-migration during the pandemic in comparison to the baseline pre-pandemic period (Investigation 2, see Section~\ref{sec:methods-macro}).

Before analyzing $v_i$, we explore how it relates to the only other known systematic quantity about distribution of family in the U.S. Namely, we check how $v_i$ relates to $\phi_i$, an estimate constructed based on a direct survey by~\citet{mcleod2023origins} of the probability that an individual living in city $i$ reports having non-household family nearby. We note that $\phi$ is not employed more broadly in this article because it is only available for 258 CBSAs, whereas our proxy $v$ is calculated for all CBSAs in the U.S. 

Following the methodology in ~\citet{mcleod2023origins}, we perform a modal regression of $v_i$ as a function of $\phi_i$. Modal regression identifies the typical behavior of a random variable as a function of some independent variable using a smoothing kernel, picking up functional relationships that traditional regressions may otherwise miss~\citep{ChenMode}. The method constructs 2-d kernel density estimates (KDE)~\citep{hastie2009elements} from our set of data points $(\phi_i, v_i)$. Then, conditioned on each value on the horizontal axis $(\phi)$, it calculates the conditional density of the KDE along the vertical axis ($v$), and extracts the local \emph{mode} of the conditional density. These local modes of $v$ are displayed as the white curve in Fig.~3 (values of KDE are visualized using a color scale). Fig.~3 shows that our parental family availability proxy $v$ is monotonically related to $\phi$, i.e. cities with larger $\phi$ also tend to have larger $v$.

\begin{figure}[t!]
    \centering
    \includegraphics[width=0.5\textwidth]{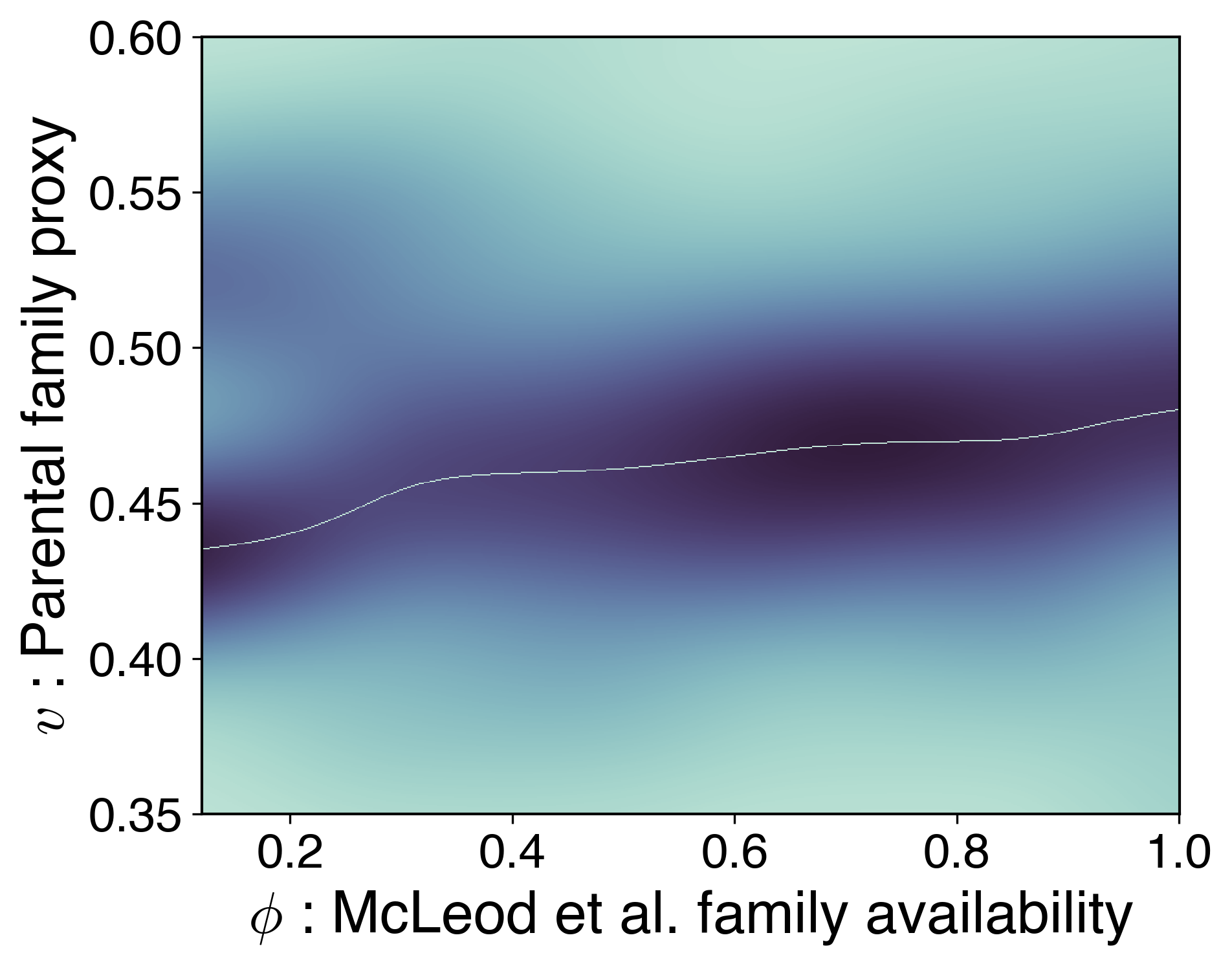}
    \caption{\color{Gray} A heatmap showing normalized density of our parental availability proxy variable $v$ conditioned on the general family availability $\phi$ obtained from~\citeauthor{mcleod2023origins}~\citeyearpar{mcleod2023origins} (see Sections~\ref{sec:results-large-scale} for discussion and methodology).}
    \label{fig:3}
\end{figure}

Now, relating $v_i$ to city population size in Fig.~1b (blue curve, left vertical axis), we find that parental family availability exhibits a decaying trend with city population (i.e., the share of households whose householders can be parents to adult children are larger in small cities than in large cities), which is consistent with the finding in~\citet{mcleod2023origins} for general family availability $\phi$.

Grouping cities by quintiles of parental family availability $v$, we see a notable increase in $y_{i}(t)$, a measure of net flux into city $i$, for those cities in the top three quintiles right after COVID-19 broke out in the U.S. when compared to the corresponding time period the year prior (Fig.~1c). A similar pattern is not seen in the bottom $v$-quintile.

\sloppy In Fig.~1b (grey line), we observe: a) an increasing relationship between $v_i$ and $\log \left[ y_{i}(\theta=1) / y_{i}(\theta=0) \right]$, and b) $\log \left[ y_{i}(\theta=1) / y_{i}(\theta=0) \right] > 0$ for cities with higher levels of parental availability. This indicates that, for those cities with larger parental family availability $v$, their net influx tended to be larger during the pandemic compared to before. (By definition, the larger $y_{i}(\theta)$ is, the larger the inflow per outflow; hence, $y_{i}(\theta=1) / y_{i}(\theta=0) > 1$ indicates that inflow per outflow was larger during the pandemic than before.)

These results are consistent with our proposition that kin partially drove the migration changes that we see at the population level. However, other population effects may be at play. For example, lower population density and cost of living may be driving people to move to smaller cities which tends to coincide with larger $v_i$. To perform a more in-depth test of the situation, we next estimate the empirical model, described in Section~\ref{sec:methods-model}, with these factors as control variables.


\subsection{Empirical model}\label{sec:results-model}

\begin{table}[ht!]
\centering
\begin{tabular}{lcccc}
\toprule
$\beta$  & (1) & (2) & (3) & (4)\\
\midrule
 $v$ (parental family availability) & 0.783*** & 0.666*** & 0.681*** & 0.631***\\
\vspace{2mm} & ({0.066}) & ({0.068}) & ({0.067}) & ({0.066})\\

\midrule
$\rho_a$  &  &  & & \\
\midrule

Population &  & {-0.018}*** & {-0.010}* & {-0.008}\\
 &  & ({0.004}) & ({0.005}) & ({0.005})\\
Population density &  & {-0.004} & {-0.007} & {-0.007}\\
 &  & ({0.006}) & ({0.006}) & ({0.006})\\
Median home value &  &  & {-0.026}* & {-0.014}\\
 &  &  & ({0.014}) & ({0.015})\\
Median income &  &  & {-0.096}*** & {-0.118}***\\
 &  &  & ({0.036}) & ({0.037})\\
Employment (jobs per person) &  &  & {0.003} & {0.003}\\
 &  &  & ({0.003}) & ({0.003})\\
Share of single-family homes &  &  &  & {0.139}**\\
 &  &  &  & ({0.062})\\
\midrule
Observations & {1852} & {1852} & {1846} & {1846}\\
$R^2$ Within & {0.138} & {0.183} & {0.216} & {0.221}\\
Adj. $R^2$ Within & {0.137} & {0.180} & {0.210} & {0.215}\\
\bottomrule
\multicolumn{5}{l}{\rule{0pt}{1em}* p $<$ 0.1, ** p $<$ 0.05, *** p $<$ 0.01. Standard errors in parentheses.}\\
\end{tabular}
\caption{\color{Gray}Empirical model (equation~(\ref{eq:regression})) results showing the positive and statistically significant effect of parental family availability on changes in the net influx into a city after the COVID-19 shock relative to before. Model (4), which includes all of the control variables, estimates that a city that is 10 percentage points higher in $v$ than another city would see a 6.5 percent higher (positive) change in inflow per outflow after the pandemic started. All models include city and time (before and after COVID-19 shock) fixed effects. All control variables except the share of SFH are in natural log scale.}
\label{table:regression}
\end{table}

We show the results of our empirical model (equation~(\ref{eq:regression})) in Table~\ref{table:regression}. In all of the models with different sets of control variables, our coefficient of interest, $\beta$, is positive and statistically significant (bold numbers in Table~\ref{table:regression}). This strongly indicates that cities with larger parental family availability, $v$, saw a higher increase in inflow per outflow. The result from Fig.~1b suggested that having higher parental family availability led a city to one of three scenarios: i) in-migration was higher during the pandemic period compared to before; or ii) out-migration was lower during the pandemic period; or iii) a combination of the two previous scenarios. The regression results suggest that these scenarios would happen to a greater extent for cities with larger $v$ compared to cities with smaller $v$. The complete model, Model (4), estimates that a city that is 10-percentage-points higher in $v$ than another city would see a 6.5 percent higher (positive) change in inflow per outflow after the pandemic started.

The coefficients of the controls in each model align with our intuition of the general relocation behavior during the pandemic. For example, we expected that both population (Models (2) and (3)) and median home value (Models (3) and (4)) would have a significant negative effect on the dependent variable as movers sought less populated and cheaper destinations. We also find that the share of SFH has a positive effect on the dependent variable.


\section{Discussion}

Overall, our empirical results support the proposition that kin ties played a role in the shift in migration to smaller cities during the COVID-19 pandemic. To the best of our knowledge, such an attempt to connect pandemic-migration to non-coresident family has not been done. Our study adds to both the migration literature and the family ties perspective by showing that while socioeconomic and physical factors such as population density and cost of living may have been at play in pandemic-migration, the picture would be incomplete if family ties are neglected (Table~\ref{table:regression}). The migrational mechanism that this study casts light upon may help in migration modeling—for example, family ties or place of nativity for subpopulations could be incorporated in models such as the generalized gravity model for human migration~\citep{park2018generalized}.

Qualitatively, if the migration decision process is thought of as ``a hierarchically ordered set of values'' or priorities~\citep{miller1976family}, our study suggests that family became more highly ranked against other factors in a systemic crisis such as the COVID-19 pandemic. The fact that we see increased out-migration from large cities and increased migration towards family or place of nativity after the pandemic started supports previous literature findings~\citep{miller1976family} that economic aspirations and extended family proximity are in tension. At the same time, this would suggest that the comparative success of large cities (see, e.g., the literature on the scaling of productivity and innovation with city size~\citep{bettencourt2007growth}) may come at a social and personal cost to individuals who have moved to these cities: they may have needed to replace relatively distant kin with local non-kin in their social network due to the cost of maintaining distant relationships, losing much of the remarkable support family provides—see also~\citet{david2019network} who explains this phenomenon.

Beyond these contributions, our study advances the emerging study of the demography-disasters nexus. As is argued by \citet{karacsonyi2021demography}, perhaps even more important than enumerating death tolls, ``the key to understanding impacts [of disasters] and avoiding them in the future is to understand the relationships between disasters and population change, both prior to and after a disaster.'' Our observations linking city population to certain trends in population realignment show how the heterogeneity in the location of extended family across the U.S. is a source of vulnerability for cities. This heterogeneity, which existed prior to the pandemic, may be due to differences in demographic, socioeconomic, or infrastructural factors. Better social or institutional support for those lacking local non-coresident family could potentially help to neutralize the effects. On the other hand, this pandemic-migration may also contribute to irreversible changes in talent availability, real estate usage, and the growth of certain industries. In this regard, future research may focus on understanding these consequences in the long run.

At the destination cities, the prioritization of face-to-face interactions with family during the lockdown stages of the pandemic suggested by survey data (see~\citet{feehan2021quantifying}) might have led to elevated transmission risks in these smaller cities and, when looked at together with other factors, may help to explain why these cities experienced comparatively worse epidemiological outcomes than in large, dense cities in later waves~\citep{cheng2020contact, koh2020we, pew-covid}. These epidemiological consequences can be long-lasting if we consider, e.g., the increased prevalence of long-COVID.

Our study is not without limitations. Most notably, part of our results rely on proxy variables of family because other large-scale data are not available that allow us to directly construct networks of movers and their family ties that also contain detailed geographic information. To this end, we combine multiple analyses at different levels in this study to provide more robust evidence of the effects of family on pandemic-migration. The lack of better data about extended family location is compatible with well-justified needs for individual privacy. At the same time, it does suggest that better sources of data that explore the spatial distribution of family across the U.S. are needed. The last systematic study of extended family across the U.S. was the now-discontinued National Survey of Families and Households; other surveys such as the Panel Study on Income Dynamics, although helpful, is not specifically designed to study non-coresident family patterns. Our results are mostly with respect to parental family, but the effects could be larger if we include other extended kin—here, again, new data would enable us to gain more insights~\citep{furstenberg2020kinship}.

To summarize, in this study we present coherent empirical evidence, using multiple sources of data, that the increased preference to migrate to smaller cities may be partly driven by the increased migration towards non-coresident family, coupled with the heterogeneous distribution of family ties in the U.S. in which people are more likely to have family ties located in smaller cities~\citep{mcleod2023origins}. On a larger scale, our study amplifies ongoing literature highlighting the role of broader kinship systems (not limited to just the nuclear family) in macro-level socioeconomic phenomena~\citep{david2019network, david2023society, furstenberg2020kinship, reed2023communication}.


\markboth{ }{ }


\newpage
\clearpage
\section*{Acknowledgements}
We thank Spectus for providing the relocation data, as well as Brennan Lake and \'{E}adaoin Ilten for interfacing on behalf of Spectus. We are also grateful to Professors Noel D. Johnson, David W. S. Wong, and Sam G. B. Roberts for their helpful comments.

\smallskip
\section*{Ethics declarations}
\subsection*{Ethical approval}
We accessed our GPS-based relocation data under a strict agreement with Spectus; the agreement precludes attempts to de-anonymize or disaggregate the data. Spectus reviewed our current study prior to journal submission.

While the IPUMS USA data are publicly available, we follow Census Bureau principles by using these data solely for statistical purposes and not attempting to disaggregate or identify any individual within the data samples.

\subsection*{Informed consent}
The Spectus data are collected from de-identified mobile-phone users who have opted in to provide access to their mobility data anonymously through a California Consumer Privacy Act (CCPA) compliant framework. The data provider, Spectus, always asks users to share their location before collecting it. Spectus also requires its application partners to disclose their relationship with Spectus and present users with options to opt out. Prior to sharing data with researchers, Spectus aggregates data to the county level. In order to further preserve privacy, Spectus discards data from counties with low thresholds of user counts.

The sampled individuals in the U.S. Census ACS are required by law to respond to the survey, but the Census Bureau is also required by law to protect respondents' privacy, including ensuring that any identifiable information is removed before publicizing the data.

\subsection*{Competing interests}
The authors declare no competing interests.

\smallskip
\section*{Authors' contributions}
Conceptualization: UK, EL;
Methodology: UK, JM, EL;
Software: UK, JM;
Investigation: UK;
Supervision: EL;
Writing- original draft: UK;
Writing- review and editing: UK, JM, EL.

\smallskip
\section*{Data availability}
The U.S. Census, IPUMS USA, Geocorr, and BEA data sets used in this study are publicly available and can be downloaded from the respective organizations' websites. The Spectus data were used under licence for the current study and are not publicly available. The code used to perform analyses of publicly available data during the current study is available at the online repository \href{https://anonymous.4open.science/r/covid_family_migration}{https://anonymous.4open.science/r/covid\_family\_migration}.

\end{document}